\documentclass[%
superscriptaddress,
preprint,
showpacs,
preprintnumbers,
 amsmath,amssymb,
 aps,
 prc,
]{revtex4-1}
\usepackage{graphicx}
\usepackage{dcolumn}
\usepackage{bm}
\usepackage{color}
\usepackage{CJK}
\def\brakket#1#2#3{\left\langle{#1}\middle|{#2}\middle|{#3}\right\rangle}
\def\ve#1{{\bm{#1}}}
\def\nuc#1#2#3{{}^{#2}_{#3}\mathrm{#1}}
\def\urm#1{\scriptstyle{\text{\textrm{\textmd{\textup{#1}}}}}}

\let\temp\epsilon
\let\epsilon\varepsilon
\let\varepsilon\temp
\let\temp\relax
\begin{document}
%
\begin{CJK*}{UTF8}{}
\preprint{RIKEN-QHP-339}
\preprint{RIKEN-iTHEMS-Report-17}
\title{Application of Coulomb energy density functional for atomic nuclei:
  Case studies of local density approximation and generalized gradient approximation}
\author{Tomoya Naito} 
\affiliation{Department of Physics, Graduate School of Science, The University of Tokyo,
  Tokyo 113-0033, Japan}
\affiliation{RIKEN Nishina Center, Wako 351-0198, Japan}
\author{Ryosuke Akashi} 
\affiliation{Department of Physics, Graduate School of Science, The University of Tokyo,
  Tokyo 113-0033, Japan}
\author{Haozhao Liang} 
\email{haozhao.liang@riken.jp}
\affiliation{RIKEN Nishina Center, Wako 351-0198, Japan}
\affiliation{Department of Physics, Graduate School of Science, The University of Tokyo,
  Tokyo 113-0033, Japan}
\date{\today}
\begin{abstract}
  We test the Coulomb exchange and correlation energy density functionals of electron systems for atomic nuclei in the local density approximation (LDA) and the generalized gradient approximation (GGA).
  For the exchange Coulomb energies, it is found that the deviation between the LDA and GGA ranges from around  $ 11 \, \% $ in $ {}^{4} \mathrm{He} $ to around $ 2.2 \, \% $ in $ {}^{208} \mathrm{Pb} $,
  by taking the Perdew-Burke-Ernzerhof (PBE) functional as an example of the GGA\@.
  For the correlation Coulomb energies, it is shown that those functionals of electron systems are not suitable for atomic nuclei.
\end{abstract}
\maketitle
\end{CJK*}
%
%
%
\section{Introduction}
\label{sec:intro}
\par
Atomic nuclei are systems that are self-bounded by the nuclear and electromagnetic forces.
Although the contribution of the nuclear force for the binding energy is much larger than that of the electromagnetic force,
it is important to study the electromagnetic contribution,
for example for the mirror nuclei mass difference \cite{Nolen1969Annu.Rev.Nucl.Sci.19_471},
the energy of the isobaric analog state \cite{Jaenecke1965Nucl.Phys.73_97,Shlomo1978Rep.Prog.Phys.41_957}, and the isospin symmetry-breaking correction to superallowed $\beta$ decay \cite{Liang2009Phys.Rev.C79_064316, Hardy2015Phys.Rev.C91_025501},
which are caused only by the electromagnetic force if the nuclear force has full isospin symmetry \cite{Wigner1937Phys.Rev.51_106}.
The form of the static electromagnetic force in the non-relativistic scheme is known as the Coulomb interaction.
Therefore, it in principle allows high-accuracy evaluation of such electromagnetic contributions.
However, so far the most widely used scheme is the Hartree-Fock or even the Hartree-Fock-Slater or Hartree approximations \cite{Bender2003Rev.Mod.Phys.75_121, Meng2006Prog.Part.Nucl.Phys.57_470, Liang2015Phys.Rep.570_1, Nakatsukasa2016Rev.Mod.Phys.88_045004}.
Moreover, the Fock potential is non-local,
and thus the corresponding numerical cost is $ O \left( N^4 \right) $,
with $ N $ the number of particles.
\par
In contrast, in electron systems, such as in atomic physics, molecular physics, and condensed matter physics,
the phenomena are intrinsically determined by the Coulomb interaction.
High-accuracy calculations for the ground-state energy have been developed,
for example by quantum Monte Carlo calculations \cite{McMillan1965Phys.Rev.138_A442,Ceperley1977Phys.Rev.B16_3081,Needs2010J.Phys.:Condens.Matter22_023201}
and density functional theory (DFT) \cite{Engel}.
It was proved by Hohenberg, Kohn, and Sham \cite{Hohenberg1964Phys.Rev.136_B864,Kohn1965Phys.Rev.140_A1133} that, in principle, DFT gives the exact ground-state energy $ E_{\urm{gs}} $ corresponding to the Hamiltonian
\begin{equation}
  H
  =
  - \frac{\hbar^2}{2m}
  \sum_j
  \nabla^2_j
  +
  \sum_j
  V_{\urm{ext}} \left( \ve{r}_j \right)
  +
  \sum_{j < k}
  V_{\urm{int}} \left( \ve{r}_j, \ve{r}_k \right)
\end{equation}
as
\begin{equation}
  E_{\urm{gs}}
  =
  T_0 \left[ \rho_{\urm{gs}} \right]
  +
  \int
  V_{\urm{ext}} \left( \ve{r} \right) \,
  \rho_{\urm{gs}} \left( \ve{r} \right) \, d \ve{r}
  +
  \frac{1}{2}
  \iint
  V_{\urm{int}} \left( \ve{r}, \ve{r}' \right) \,
  \rho_{\urm{gs}} \left( \ve{r} \right) \,
  \rho_{\urm{gs}} \left( \ve{r}' \right)
  \, d \ve{r} \,d \ve{r}'
  +
  E_{\urm{xc}} \left[ \rho_{\urm{gs}} \right],
\end{equation}
where $ \rho_{\urm{gs}} $ is the ground-state density distribution,
$ T_0 $ is the kinetic energy of the non-interacting reference system,
$ m $ is the mass,
and $ V_{\urm{ext}} $ and $ V_{\urm{int}} $ are the external field and two-body interaction, respectively.
The exchange-correlation energy density functional (EDF) $ E_{\urm{xc}} \left[ \rho \right] $ includes the correction of kinetic energy for the interacting system
from the non-interacting reference system \cite{Hohenberg1964Phys.Rev.136_B864,Kohn1965Phys.Rev.140_A1133}.
The accuracy of DFT depends only on the accuracy of the exchange-correlation EDF\@.
High-accuracy non-empirical exchange-correlation EDFs for electron systems have been proposed for decades
\cite{Vosko1980Can.J.Phys.58_1200--1211,Perdew1981Phys.Rev.B23_5048--5079,Perdew1992Phys.Rev.B45_13244--13249,Becke1988Phys.Rev.A38_3098,Perdew1992Phys.Rev.B46_6671,Perdew1996Phys.Rev.Lett.77_3865,Perdew2008Phys.Rev.Lett.100_136406,Lee1988Phys.Rev.B37_785--789},
although a systematic way of deriving the exact EDF is still an open problem \cite{Perdew2001AIPConf.Proc.577_1,Drut2010Prog.Part.Nucl.Phys.64_120,Liang2018Phys.Lett.B779_436}.
The numerical cost of DFT calculation is $ O \left( N^3 \right) $,
and high-accuracy large-scale calculation is thus easier to perform than other methods with similar accuracy.
\par
From the point of view of the electromagnetic force, protons in atomic nuclei and electrons in electron systems share common properties except for the difference in mass and the sign of the charge.
Therefore, it is interesting to investigate to what extent the knowledge of electron systems is applicable for studying the effect of the electromagnetic force in atomic nuclei.
\par
In this paper, we test the exchange and correlation EDFs of electron systems in the context of atomic nuclei.
Both the local density approximation (LDA) and generalized gradient approximation (GGA) functionals are investigated.
The error due to the approximations in the EDF is separable into two parts: density-driven error and functional-driven error \cite{Kim2013Phys.Rev.Lett.111_073003}.
In this work, we use the experimentally observed charge-density distribution for quantitative calculations of selected nuclei to avoid the first error.
For the second part, a straightforward application of the EDFs developed for electron systems obviously suffers from the subtle errors due to the coexistence of the Coulomb and nuclear forces.
Nevertheless, in atomic nuclei, the effects of different many-body correlations are not cleanly isolated, when fitting new nuclear EDFs \cite{Goriely2008Phys.Rev.C77_031301,Kortelainen2012Phys.Rev.C85_024304}.
Also, see the discussion in Sec.~\ref{sec:electron_sep} below.
\par
This paper is organized in following way:
First, in Sec.~\ref{sec:electron} we
show the general expressions of LDA and GGA functionals
and discuss the separability of exchange and correlation functionals.
Then in Sec.~\ref{sec:result} we show the calculations by using experimental charge-density distributions.
Finally, in Sec.~\ref{sec:conc} we give the conclusion and perspectives.
In the Appendix, we show the details of Coulomb EDFs.
%

%
%
%
\section{Exchange and Correlation Energy Density Functionals}
\label{sec:electron}
\subsection{General Expressions}
\par
It is assumed the exchange-correlation EDF is divided into two parts,
the exchange EDF $ E_{\urm{x}} \left[ \rho \right] $ and the correlation EDF $ E_{\urm{c}} \left[ \rho \right] $ as \cite{Engel}
\begin{equation}
  E_{\urm{xc}} \left[ \rho \right]
  =
  E_{\urm{x}} \left[ \rho \right]
  +
  E_{\urm{c}} \left[ \rho \right],
\end{equation}
and both EDFs are written as
\begin{equation}
  E_i \left[ \rho \right]
  =
  \int
  \epsilon_i \left[ \rho \right] \,
  \rho \left( \ve{r} \right) \,
  d \ve{r}
  \qquad
  \text{($ i = \mathrm{x}, \, \mathrm{c} $)},
\end{equation}
where  $ \epsilon_i\left[ \rho \right] $ as a functional of density is called the energy density in electron systems, which corresponds to the concept of energy per particle in nuclear physics.
\par
When it is assumed that the energy density depends only on the density at $ \ve{r} $ locally as
\begin{equation}
  E_i \left[ \rho \right]
  =
  \int
  \epsilon_i^{\urm{LDA}} \left( \rho \left( \ve{r} \right) \right) \,
  \rho \left( \ve{r} \right) \,
  d \ve{r}
  \qquad
  \text{($ i = \mathrm{x}, \, \mathrm{c} $)},
\end{equation}
this approximation is called the LDA\@.
The LDA gives the exact solutions for the systems with homogeneous density distribution,
and it also gives high-accuracy results for the systems with nearly constant density distribution.
\par
In the GGA, the energy density depends not only on the density distribution $ \rho $ but also on its gradient $ \left| \nabla \rho \right| $ at $ \ve{r} $ locally.
It is expressed as
\begin{equation}
  E_i \left[ \rho \right]
  =
  \int
  \epsilon_i^{\urm{GGA}} \left( \rho \left( \ve{r} \right), \left| \ve{\nabla} \rho \left( \ve{r} \right) \right| \right) \,
  \rho \left( \ve{r} \right) \,
  d \ve{r}
  \qquad
  \text{($ i = \mathrm{x}, \, \mathrm{c} $)}.
\end{equation}
Several non-empirical GGA functionals have been proposed \cite{Becke1988Phys.Rev.A38_3098, Perdew1992Phys.Rev.B46_6671, Perdew1996Phys.Rev.Lett.77_3865, Perdew2008Phys.Rev.Lett.100_136406}.
\par
See the Appendix for the details of Coulomb EDFs as well as the translation from the Hartree atomic unit to the general unit.
\subsection{Separability of Exchange and Correlation Functionals}
\label{sec:electron_sep}
\par
In the following, let us discuss carefully the separability of exchange and correlation functionals.
According to the Hohenberg-Kohn theorem~\cite{Hohenberg1964Phys.Rev.136_B864},
the ground-state energy is written as
\begin{equation}
  \label{eq:HK}
  E_{\urm{gs}}
  =
  F \left[ \rho_{\urm{gs}} \right]
  +
  \int
  \rho_{\urm{gs}}
  \left( \ve{r} \right) \,
  V_{\urm{ext}}
  \left( \ve{r} \right) \,
  d \ve{r},
\end{equation}
with the universal functional $ F $ for each given interaction $ V_{\urm{int}} $.
The universal functional is written as
\begin{equation}
  F \left[ \rho \right]
  =
  T_0 \left[ \rho \right]
  +
  \frac{1}{2}
  \iint
  V_{\urm{int}} \left( \ve{r}, \ve{r}' \right) \,
  \rho \left( \ve{r} \right) \,
  \rho \left( \ve{r}' \right)
  \, d \ve{r} \,d \ve{r}'
  +
  E_{\urm{x}} \left[ \rho \right]
  +
  E_{\urm{c}} \left[ \rho \right],
\end{equation}
in the Kohn-Sham scheme \cite{Kohn1965Phys.Rev.140_A1133}.
\par
Here, we consider two systems: System 1 has one interaction $ V_1 $, and system 2 has two interactions, $ V_1 $ and $ V_2 $.
We also define the universal functional $ F_1 $ for the interaction $ V_1 $ and the functional $ F_{1+2} $ for the interaction $ V_1 + V_2 $.
The universal functionals $ F_1 $ and $ F_{1+2} $ correspond to
\cite{Levy1979Proc.Natl.Acad.Sci.USA76_6062,Levy1982Phys.Rev.A26_1200}
\begin{align}
  F_1 \left[ \rho \right]
  & =
    \inf_{\Psi \in \mathcal{W}_{\rho}}
    \left[
    \brakket{\Psi}{T}{\Psi}
    +
    \brakket{\Psi}{V_1}{\Psi}
    \right] \notag \\
  & =
    T_0 \left[ \rho \right]
    +
    \frac{1}{2}
    \iint
    V_1 \left( \ve{r}, \ve{r}' \right) \,
    \rho \left( \ve{r} \right) \,
    \rho \left( \ve{r}' \right)
    \, d \ve{r} \,d \ve{r}'
    +
    E_{\urm{x}}^1 \left[ \rho \right]
    +
    E_{\urm{c}}^1 \left[ \rho \right],
    \label{eq:F1} \\
  F_{1+2} \left[ \rho \right]
  & =
    \inf_{\Psi \in \mathcal{W}_{\rho}}
    \left[
    \brakket{\Psi}{T}{\Psi}
    +
    \brakket{\Psi}{V_1}{\Psi}
    +
    \brakket{\Psi}{V_2}{\Psi}
    \right] \notag \\
  & =
    T_0 \left[ \rho \right]
    +
    \frac{1}{2}
    \iint
    \left[
    V_1 \left( \ve{r}, \ve{r}' \right)
    +
    V_2 \left( \ve{r}, \ve{r}' \right)
    \right]
    \rho \left( \ve{r} \right) \,
    \rho \left( \ve{r}' \right)
    \, d \ve{r} \,d \ve{r}'
    +
    E_{\urm{x}}^{1+2} \left[ \rho \right]
    +
    E_{\urm{c}}^{1+2} \left[ \rho \right],
    \label{eq:F2}
\end{align}
where $ T $ is the kinetic operator, and $ \mathcal{W}_{\rho} $ is the set of the $ N $-particle wave functions $ \Psi $
which satisfy
\begin{equation}
  \rho \left( \ve{r} \right)
  =
  N
  \int
  \Psi^* \left( \ve{r}, \ve{r}_2 , \ve{r}_3 , \ldots , \ve{r}_N \right) \,
  \Psi \left( \ve{r}, \ve{r}_2 , \ve{r}_3 , \ldots , \ve{r}_N \right) \,
  d \ve{r}_2 \, d \ve{r}_3 \, \cdots \, d \ve{r}_N.
\end{equation}
\par
In order to compare with $F_1$, we define $ F_{1+2}^1 $ and $ F_{1+2}^2 $ as
\begin{align}
  F_{1+2}^1 \left[ \rho \right]
  & =
    \brakket{\Psi_0}{T}{\Psi_0}
    +
    \brakket{\Psi_0}{V_1}{\Psi_0} \notag \\
  & =
    T_0 \left[ \rho \right]
    +
    \frac{1}{2}
    \iint
    V_1 \left( \ve{r}, \ve{r}' \right) \,
    \rho \left( \ve{r} \right) \,
    \rho \left( \ve{r}' \right)
    \, d \ve{r} \,d \ve{r}'
    +
    \tilde{E}_{\urm{x}}^1 \left[ \rho \right]
    +
    \tilde{E}_{\urm{c}}^1 \left[ \rho \right], \label{eq:F12}\\
  F_{1+2}^2 \left[ \rho \right]
  & =
    \brakket{\Psi_0}{V_2}{\Psi_0} \notag \\
  & =
    \frac{1}{2}
    \iint
    V_2 \left( \ve{r}, \ve{r}' \right) \,
    \rho \left( \ve{r} \right) \,
    \rho \left( \ve{r}' \right)
    \, d \ve{r} \,d \ve{r}'
    +
    \tilde{E}_{\urm{x}}^2 \left[ \rho \right]
    +
    \tilde{E}_{\urm{c}}^2 \left[ \rho \right],
\end{align}
where $ \Psi_0 $ gives the infimum value of Eq.~\eqref{eq:F2}.
Because of the variational principle, the following inequality holds
\begin{equation}
  \label{eq:ineq}
  F_{1+2}^1
  \left[ \rho \right]
  \ge
  F_1
  \left[ \rho \right].
\end{equation}
Therefore, there is no guarantee to assume the same exchange-correlation EDFs in both system 1 and system 2.
\par
For the exchange EDF, if the Fock term is defined as the exchange term of the Kohn-Sham orbitals, the Fock term of two interactions $ V_1 $ and $ V_2 $ in system 2 are separable as
\begin{equation}
  E_{\urm{F}}^{1+2}
  =
  E_{\urm{F}}^1 + E_{\urm{F}}^2,
\end{equation}
where $ E_{\urm{F}}^i $ denotes the Fock term for interaction, $ V_i $ ($ i = 1, \, 2 $, and $ 1+2 $).
In the homogeneous systems, the exchange EDF $ E^i_{\urm{x}} \left[ \rho \right] $ is identical to the Fock term $ E_{\urm{F}}^i $.
As a result, $ \tilde{E}_{\urm{x}}^1 \left[ \rho \right] $ in Eq.~\eqref{eq:F12} is equal to $ E_{\urm{x}}^1 \left[ \rho \right] $ in Eq.~\eqref{eq:F1}.
Note that, in inhomogeneous systems, $ E^i_{\urm{F}} $ and $ E^i_{\urm{x}} \left[ \rho \right] $ are different.
As a result, $ \tilde{E}_{\urm{x}}^1 \left[ \rho \right] $ is only approximately equal to $ E_{\urm{x}}^1 \left[ \rho \right] $ in the GGA\@.
\par
For the correlation EDF,
the difference between $ E_{\urm{c}}^1 \left[ \rho \right] $ and $ \tilde{E}_{\urm{c}}^1 \left[ \rho \right] $ includes the difference of $ F_{1+2}^1 - F_1 $ as shown in Eq.~\eqref{eq:ineq}.
In addition, the correlation EDF includes the deviation between the kinetic energy of the realistic interacting system and that of the non-interacting reference system.
Such a deviation is caused by all interactions.
Therefore, the separability of the correlation EDF is in question.
\par
In the present context, system 1 and 2 correspond to electron systems and atomic nuclei, respectively.
We apply the EDFs of electron systems to atomic nuclei as a test of these functionals, by keeping the above discussions in mind.
\par
In addition, we note that in nuclear DFT, the Coulomb part of the EDF does not include the correlation energy, and the nuclear part of the EDF is determined by parameter fittings correspondingly.
Thus, the nuclear part includes all the correlation effects, including the Coulomb one.
If the Coulomb correlation EDF is considered while the nuclear part remains the same, a part of the correlation is double counted.
In such a case, it is necessary to refit the parameters in the nuclear part of the EDFs accordingly.
Nevertheless, the ultimate refitted nuclear EDFs can reproduce not only nuclear masses, radii, etc.,~but also the experimental charge-density distributions.
Therefore, the issue of refitting nuclear EDFs will not matter in the following results and conclusions,
which are calculated directly from the experimental charge-density distributions $ \rho_{\urm{ch}} $.

%
%
%
\section{Results and Discussion}
\label{sec:result}
\par
In this section, the Coulomb exchange and correlation functionals $ E_{\urm{x}} \left[ \rho \right] $ and $ E_{\urm{c}} \left[ \rho \right] $ in LDA and GGA are applied to atomic nuclei.
Different versions of $ \epsilon_{\urm{c}}^{\urm{LDA}} $ \cite{Vosko1980Can.J.Phys.58_1200--1211,Perdew1981Phys.Rev.B23_5048--5079,Perdew1992Phys.Rev.B45_13244--13249} behave almost the same in the density region of nuclei,
and therefore we focus on the results obtained with PZ81 \cite{Perdew1981Phys.Rev.B23_5048--5079}, and denote them as LDA hereafter.
Nevertheless, as discussed at the end of this section, the Coulomb correlation functionals $ E_{\urm{c}} \left[ \rho \right] $ are not suitable for atomic nuclei,
and thus the GGA Coulomb correlation functionals are not discussed explicitly.
\subsection{Calculations for $ \nuc{Pb}{208}{} $}
\begin{figure}
  \centering
  \includegraphics[width=8cm]{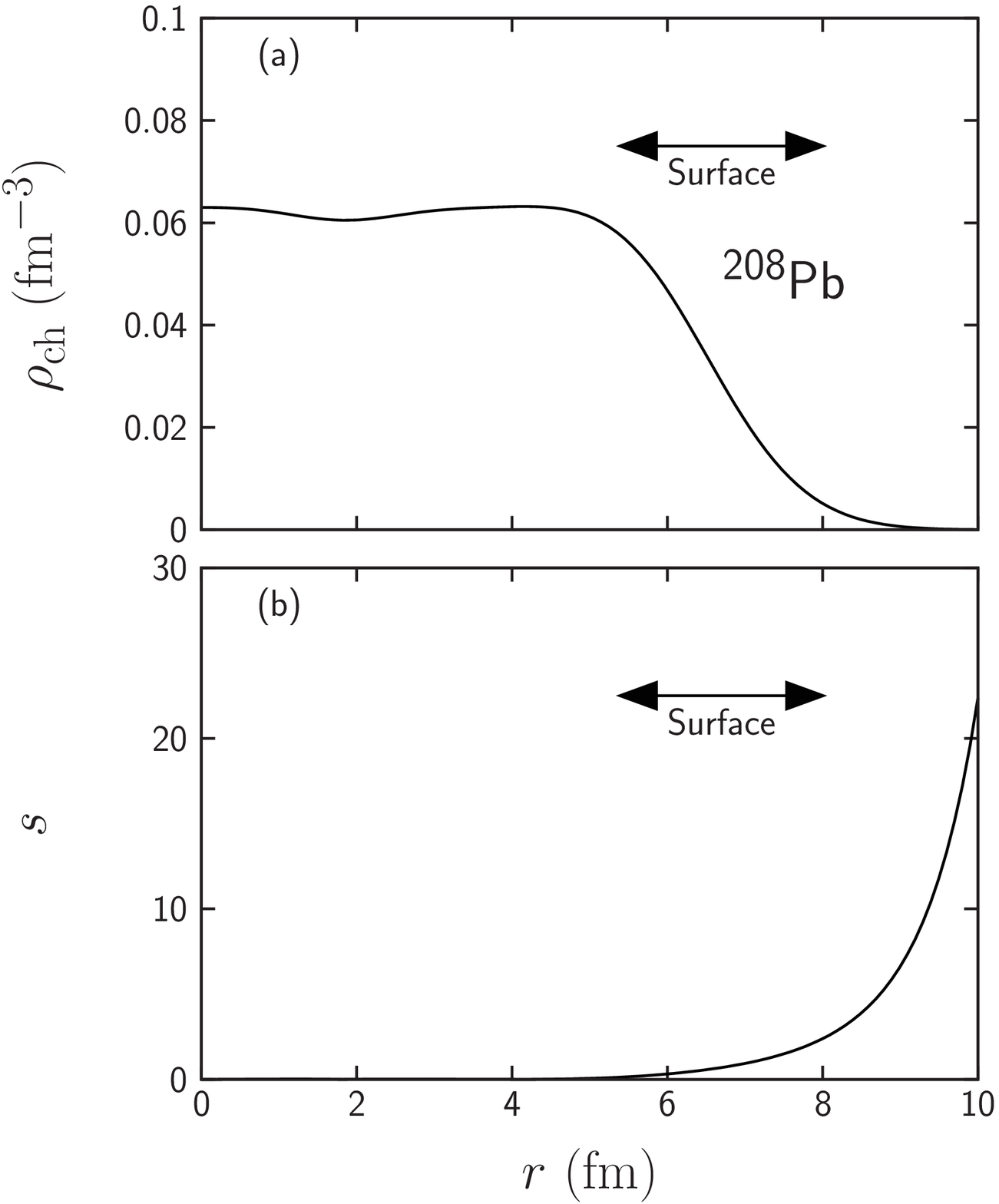}
  \caption{(a) Experimental charge-density distribution $ \rho_{\urm{ch}} \left( r \right) $ for $ \nuc{Pb}{208}{} $ \cite{Vries1987At.DataNucl.DataTables36_495}.
  The surface is defined as the region where the density is between $ 90 \, \%$ and $ 10 \, \% $ of the maximum density.
  (b) Dimensionless density gradient $ s $ as a function of $ r $.}
  \label{fig:dens_pb}
\end{figure}
\begin{figure}
  \centering
  \includegraphics[width=8cm]{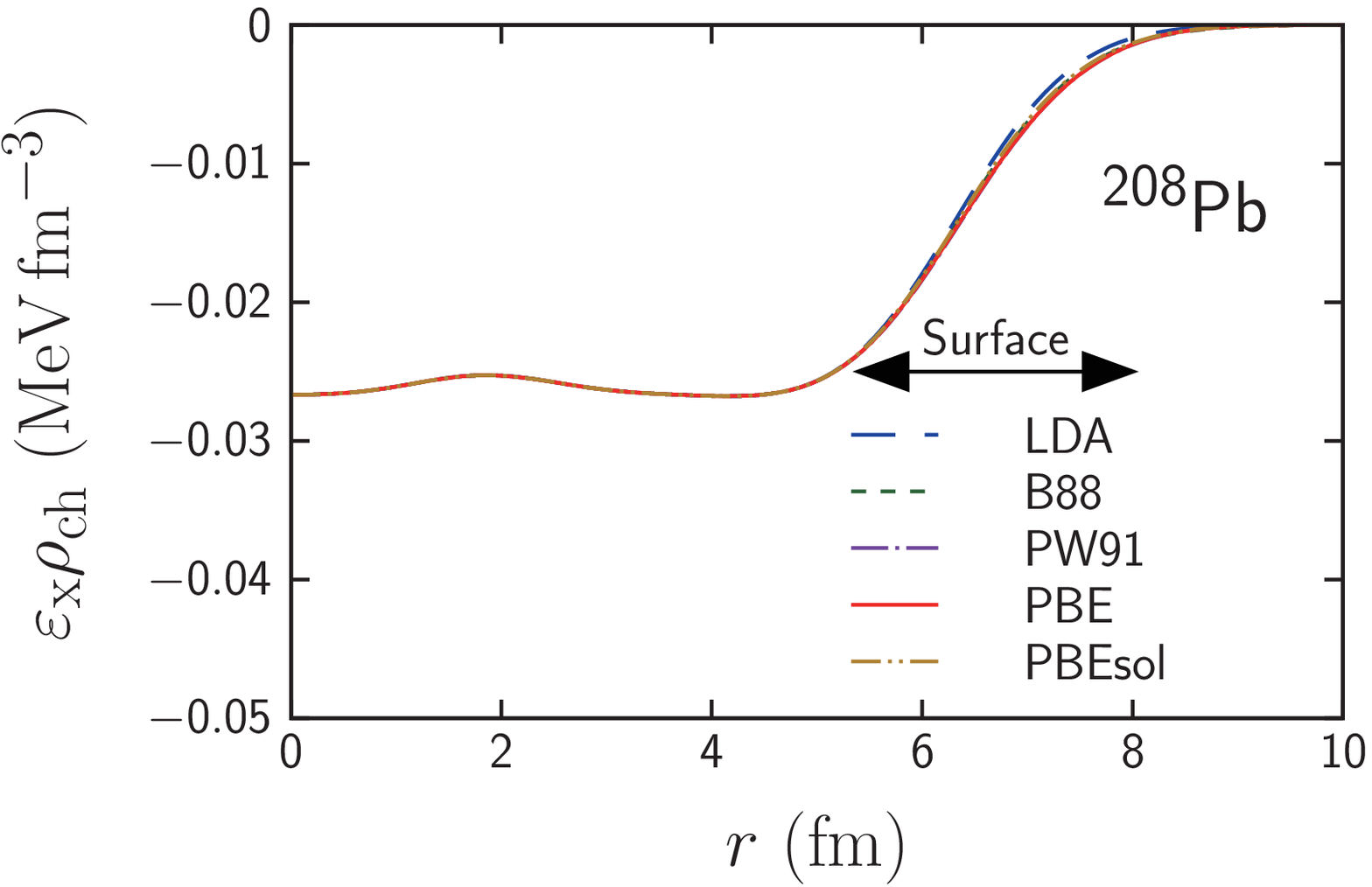}
  \caption{GGA exchange energy densities weighted with $ \rho_{\urm{ch}} $ for $ \nuc{Pb}{208}{} $ as a function of $ r $.
    The LDA result is shown with the long-dashed line.
    Those given by the GGA functionals B88 \cite{Becke1988Phys.Rev.A38_3098}, PW91 \cite{Perdew1992Phys.Rev.B46_6671}, PBE \cite{Perdew1996Phys.Rev.Lett.77_3865}, and PBEsol \cite{Perdew2008Phys.Rev.Lett.100_136406} are shown with the short-dashed, dot-dashed, solid, and dot-dot-dashed lines, respectively.}
  \label{fig:gga_pb}
\end{figure}
\par
In the following discussions, we use the experimental charge-density distributions $ \rho_{\urm{ch}} \left( r \right) $ given by the sum-of-Gaussian analysis in Ref.~\cite{Vries1987At.DataNucl.DataTables36_495} as the inputs of ground-state density distributions for testing the LDA and GGA exchange and correlation functionals.
\par
The charge-density distribution $ \rho_{\urm{ch}}\left( r \right) $ of $ \nuc{Pb}{208}{} $ is shown in Fig.~\ref{fig:dens_pb}(a).
The surface is defined as the region where the density is between $ 90 \, \%$ and $ 10 \, \% $ of the maximum density.
A dimensionless density gradient $ s $
and the local Fermi wave number $ k_{\urm{F}} $ are given as
\begin{equation}
  \label{eq:GGA_s}
  s
  =
  \frac{\left| \nabla \rho \right|}{2 k_{\urm{F}} \rho},
  \qquad
  k_{\urm{F}}
  =
  \left(3 \pi^2 \rho \right)^{1/3},
\end{equation}
respectively.
The corresponding dimensionless density gradient $ s $ is shown as a function of $ r $ in Fig.~\ref{fig:dens_pb}(b).
The corresponding GGA exchange energy density weighted with $ \rho_{\urm{ch}}\left( r \right) $ for $ \nuc{Pb}{208}{} $ is shown in Fig.~\ref{fig:gga_pb}.
On the one hand, the LDA result is shown with the long-dashed line.
On the other hand, those given by the GGA functionals
B88 \cite{Becke1988Phys.Rev.A38_3098}, PW91 \cite{Perdew1992Phys.Rev.B46_6671}, PBE \cite{Perdew1996Phys.Rev.Lett.77_3865}, and PBEsol \cite{Perdew2008Phys.Rev.Lett.100_136406}
are shown with the short-dashed, dot-dashed, solid, and dot-dot-dashed lines, respectively.
\par
In the central region $ r \lesssim 5 \, \mathrm{fm} $, the density is almost constant with a value around half of the saturation density, and thus the dimensionless density gradient $ s $ is almost equal to zero.
Therefore, the LDA and GGA give almost the same $ \epsilon_{\urm{x}} \rho_{\urm{ch}} $.
In contrast, $ s $ increases substantially with $ r $ outside the central region.
In particular, in the surface region, corresponding to $ 5.4 \lesssim r \lesssim 8.0 \, \mathrm{fm} $,
the dimensionless density gradient $ s $ reads
$ 0.14 \lesssim s \lesssim 2.0 $.
It is seen that the $ \epsilon_{\urm{x}} \rho_{\urm{ch}} $ given by the LDA and GGA diverge from each other, while those given by different GGA functionals are quite similar.
Outside of the surface region, $ s $ keeps increasing, but it is not relevant to the $ E_{\urm{x}} $ since the charge-density distribution $ \rho_{\urm{ch}}\left( r \right) $ decreases exponentially.
\par
The Coulomb energy is calculated separately as the direct term $ E_{\urm{d}} $, exchange term $ E_{\urm{x}} $, and correlation term $ E_{\urm{c}} $.
The direct term reads
\begin{equation}
  E_{\urm{d}}
  =
  \frac{1}{2}
  \frac{e^2}{4 \pi \epsilon_0}
  \iint
  \frac{\rho_{\urm{ch}} \left( \ve{r} \right) \, \rho_{\urm{ch}} \left( \ve{r}' \right)}
  {\left| \ve{r} - \ve{r}' \right|}
  \, d \ve{r} \, d \ve{r}'.
\end{equation}
In the level of LDA, the results for $ \nuc{Pb}{208}{} $ are shown in Table~\ref{tab:lda}.
It is found that the ratio of the correlation energy to the exchange energy is around $ 1.8 \, \% $, which is consistent with the estimate of order of magnitude in Fig.~\ref{fig:ratio} shown in the Appendix.
In the level of GGA, the results of the exchange Coulomb energy are shown in Table~\ref{tab:gga_x},
where four different GGA exchange functionals are used.
\par
Here $ \Delta E_{\urm{x}} $ denotes the deviations between the LDA and GGA in the exchange energy as
\begin{equation}
  \label{eq:delta_x} 
  \Delta E_{\urm{x}}
  =
  \frac{E_{\urm{x}}^{\urm{GGA}} - E_{\urm{x}}^{\urm{LDA}}}
  {E_{\urm{x}}^{\urm{GGA}}}.
\end{equation}
For the GGA exchange energy $ E_{\urm{x}} $, an overall enhancement around $ 2 \, \% $ is found compared to the LDA one.
Among different functionals, the PW91 and PBE show the largest enhancements, $ \Delta E_{\urm{x}}^{\urm{PW91}} \simeq 2.3 \, \% $ and $ \Delta E_{\urm{x}}^{\urm{PBE}} \simeq 2.2 \, \% $, respectively, whereas the PBEsol shows the smallest enhancement, $ \Delta E_{\urm{x}}^{\urm{PBEsol}} \simeq  1.4 \, \% $.
This indicates the differences between the GGA and LDA exchange energies are around $ 500 \, \mathrm{keV} $, which are not negligible for the discussions of Coulomb energy or nuclear mass.
The main enhancement comes from the surface region, where the enhancement factor reaches around $ 1.5 $ as shown in Fig.~\ref{fig:gga-f}.
Thus, it is expected that for lighter nuclei the overall enhancement will further increase.
\par
\subsection{Calculations for $ \nuc{He}{4}{} $ and $ \nuc{O}{16}{} $}
\begin{figure}
  \centering
  \includegraphics[width=8cm]{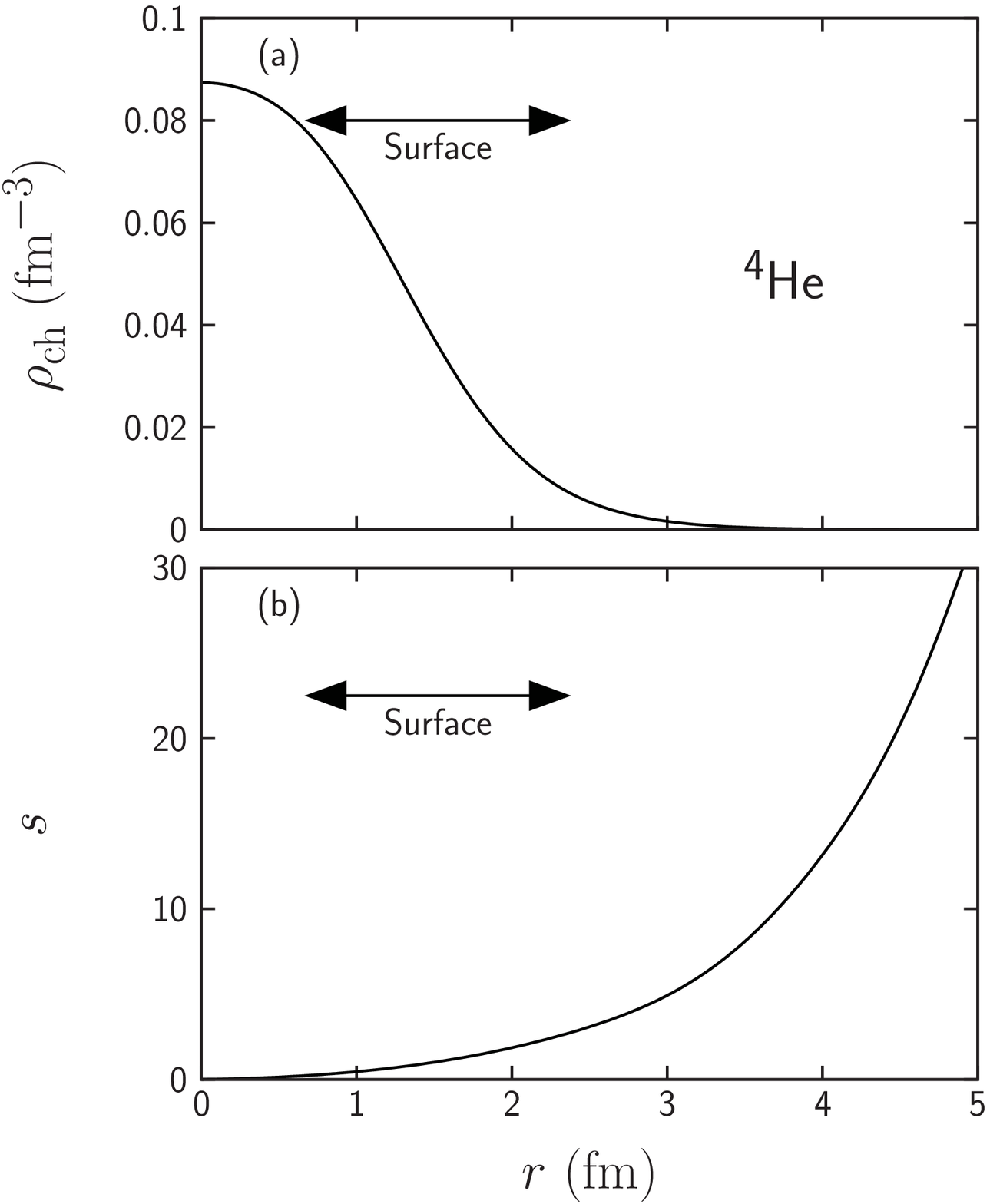}
  \caption{Same as Fig.~\ref{fig:dens_pb} but for $ \nuc{He}{4}{} $.}
  \label{fig:dens_he}
\end{figure}
\begin{figure}
  \centering
  \includegraphics[width=8cm]{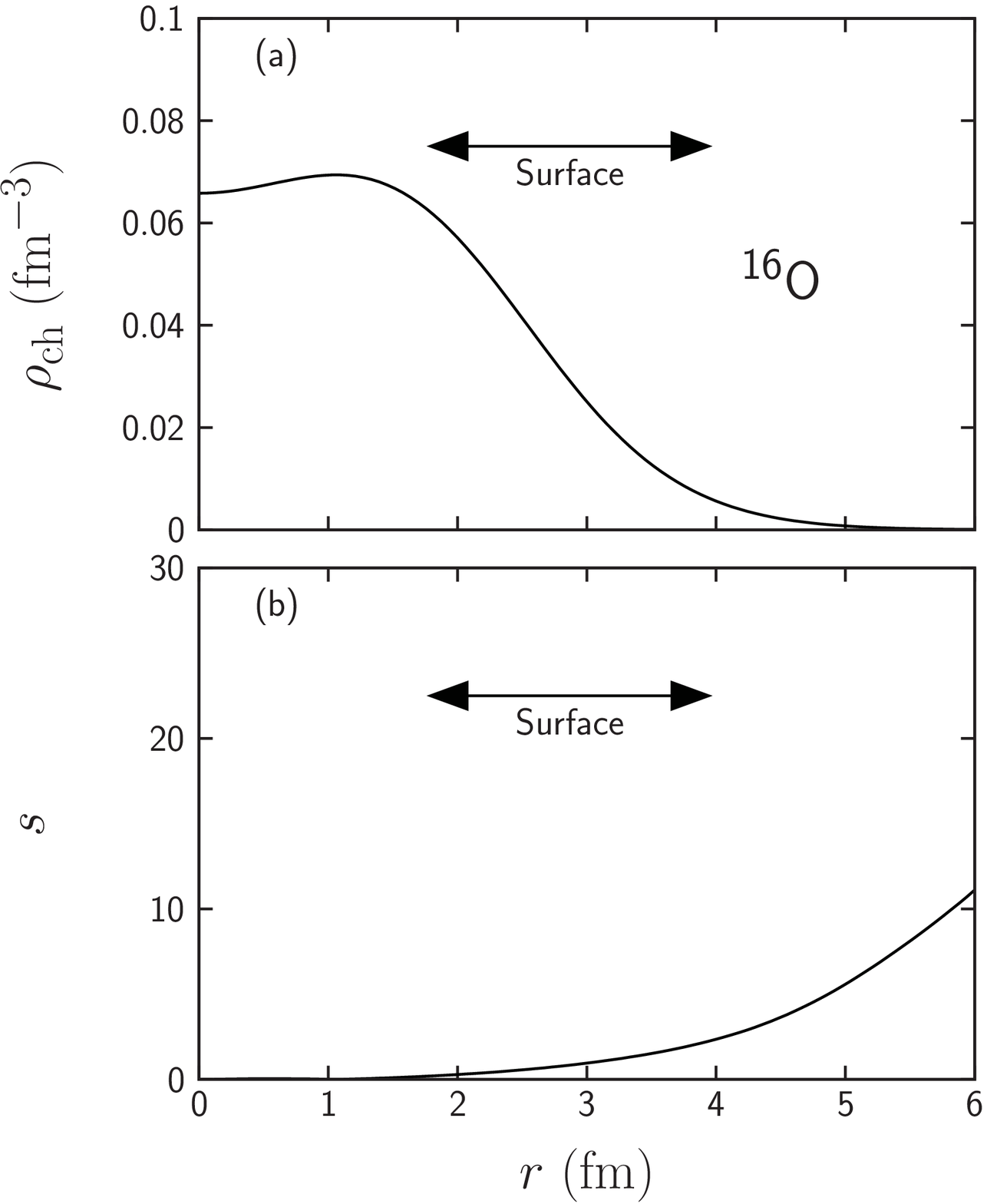}
  \caption{Same as Fig.~\ref{fig:dens_pb} but for $ \nuc{O}{16}{} $.}
  \label{fig:dens_o}
\end{figure}
\begin{figure}
  \centering
  \includegraphics[width=8cm]{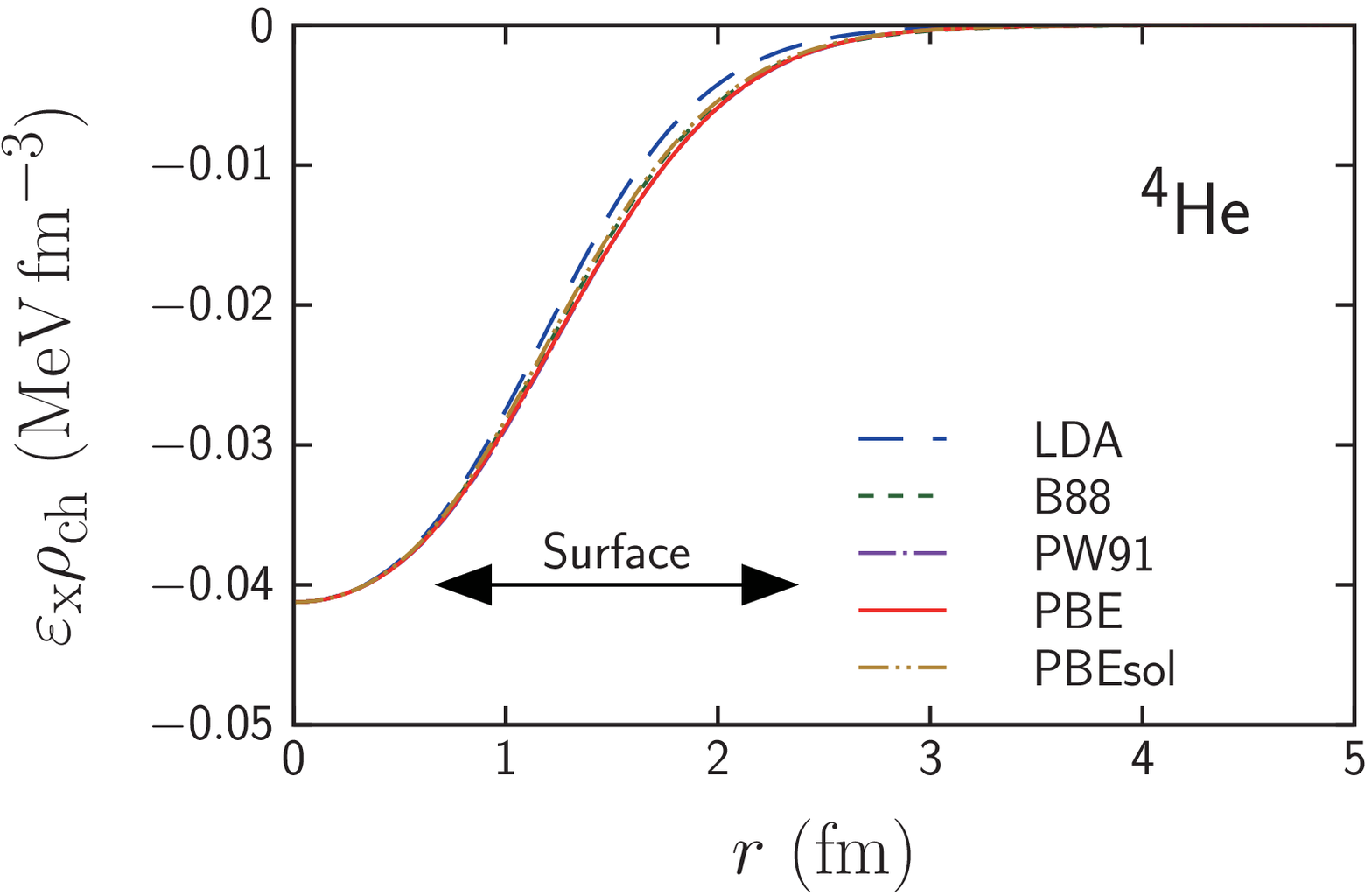}
  \caption{Same as Fig.~\ref{fig:gga_pb} but for $ \nuc{He}{4}{} $.}
  \label{fig:gga_he}
\end{figure}
\begin{figure}
  \centering
  \includegraphics[width=8cm]{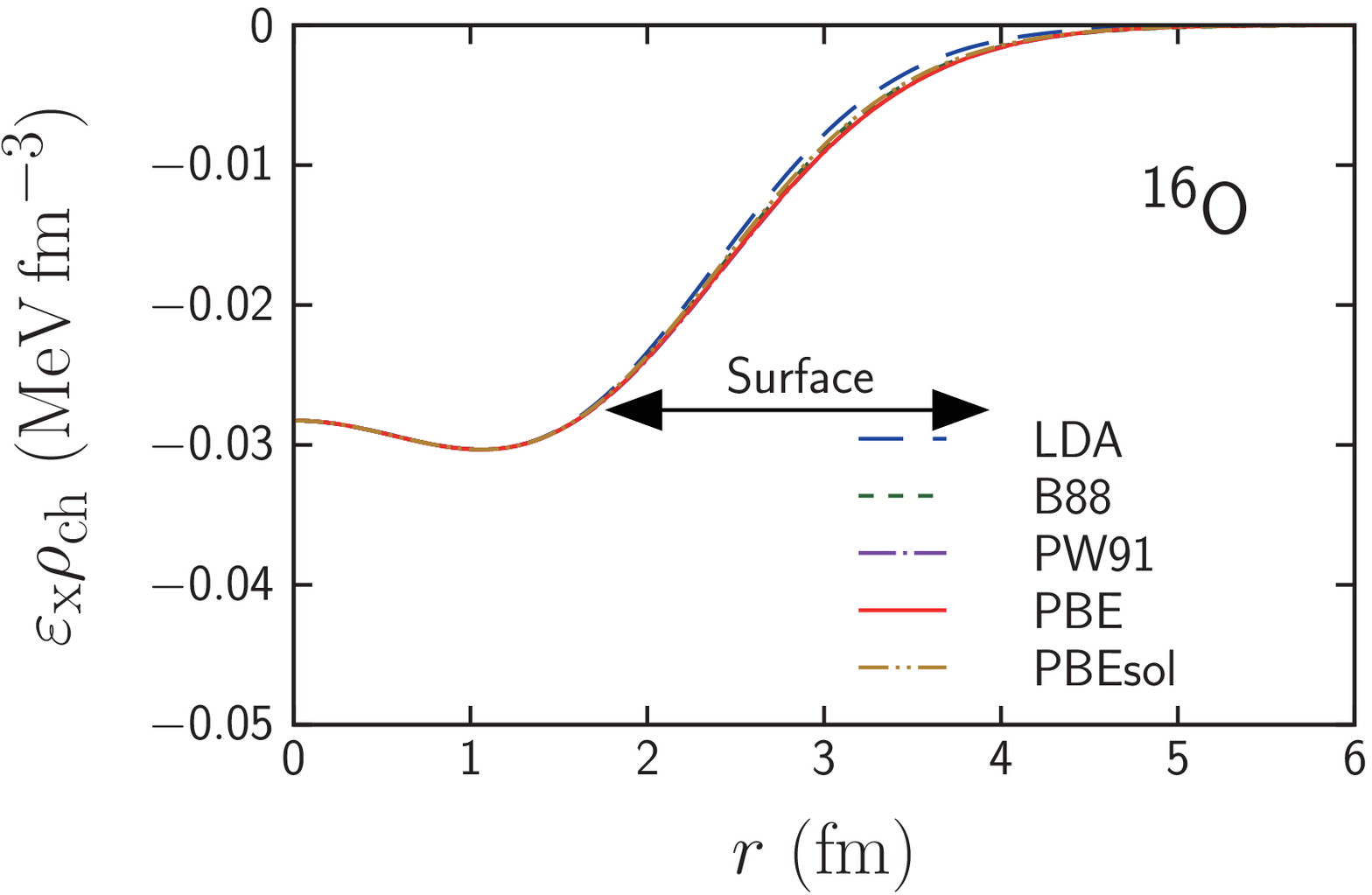}
  \caption{Same as Fig.~\ref{fig:gga_pb} but for $ \nuc{O}{16}{} $.}
  \label{fig:gga_o}
\end{figure}
\par
In order to see the effects of the surface region, here we compare the results of the heavy nucleus $ \nuc{Pb}{208}{} $ to the light nuclei $ \nuc{He}{4}{} $ and $ \nuc{O}{16}{} $.
The experimental charge-density distributions $ \rho_{\urm{ch}} \left( r \right) $ and the corresponding dimensionless density gradients $ s $ for $ \nuc{He}{4}{} $ and $ \nuc{O}{16}{} $ are shown in Figs.~\ref{fig:dens_he} and \ref{fig:dens_o}, respectively.
It is seen that the behaviors of $ \rho_{\urm{ch}} $ and $ s $ for $ \nuc{He}{4}{} $ and $ \nuc{O}{16}{} $ are almost the same as those for $ \nuc{Pb}{208}{} $, while the ratio of the surface region to the whole nuclear region is larger for light nuclei.
Note that the ratio $ \epsilon_{\urm{c}}/\epsilon_{\urm{x}} $ increases as $ \rho_{\urm{ch}} $ decreases as shown in Fig.~\ref{fig:ratio}.
Therefore, the ratio $ E_{\urm{c}}/E_{\urm{x}}$ for light nuclei is enhanced slightly compared with that for heavier nuclei (see Table~\ref{tab:lda}).
\par
In Figs.~\ref{fig:gga_he} and \ref{fig:gga_o}, the GGA exchange energy density weighted with $ \rho_{\urm{ch}}\left( r \right) $ is shown for $ \nuc{He}{4}{} $ and $ \nuc{O}{16}{} $, respectively.
Comparing with those shown in Fig.~\ref{fig:gga_pb}, the difference between the LDA and GGA are more visible by using the same scale.
Therefore, by taking the PBE functional as an example, the GGA exchange energies show enhancements of $ \Delta E_{\urm{x}}^{\urm{PBE}} \simeq  11 \, \% $ and $ \Delta E_{\urm{x}}^{\urm{PBE}} \simeq  6.0 \, \% $ for $ \nuc{He}{4}{} $ and $ \nuc{O}{16}{} $, respectively.
\subsection{Calculations from Light to Heavy Nuclei}
\begin{table}[!htb]
  \centering
  \caption{Direct $E_{\urm{d}}$, exchange $E_{\urm{x}}$, and correlation $E_{\urm{c}}$ Coulomb energies in the LDA for selected nuclei, together with the ratios of $ E_{\urm{c}} /E_{\urm{x}} $.
    The experimental root-mean-square charge radii $ \left\langle r_{\urm{ch}}^2 \right\rangle^{1/2} $
    \cite{Vries1987At.DataNucl.DataTables36_495} are also shown.}
  \label{tab:lda}
  \begin{tabular}{rddddd}
    \hline \hline
    Nuclei & \multicolumn{1}{c}{$ \left\langle r_{\urm{ch}}^2 \right\rangle^{1/2} $ ($ \mathrm{fm} $)} & \multicolumn{1}{c}{$E_{\urm{d}}$ ($ \mathrm{MeV} $)} & \multicolumn{1}{c}{$E_{\urm{x}}$ ($ \mathrm{MeV} $)} & \multicolumn{1}{c}{$E_{\urm{c}}$ ($ \mathrm{MeV} $)} & \multicolumn{1}{c}{$ E_{\urm{c}} /E_{\urm{x}} $ ($ \% $)} \\ \hline
    $ \nuc{He}{4}{} $ & 1.676 (8) & 1.518 & -0.6494 & -0.01296 & 1.996 \\
    $ \nuc{C}{12}{} $ & 2.469 (6) & 9.481 & -1.962 & -0.03904 & 1.990 \\
    $ \nuc{O}{16}{} $ & 2.711 & 15.41 & -2.638 & -0.05218 & 1.978 \\
    $ \nuc{Ca}{40}{} $ & 3.480 (3) & 75.74 & -7.087 & -0.1329 & 1.875 \\
    $ \nuc{Ca}{48}{} $ & 3.460 & 75.68 & -7.113 & -0.1332 & 1.873 \\
    $ \nuc{Ni}{58}{} $ & 3.772 (4) & 136.6 & -10.28 & -0.1879 & 1.828 \\
    $ \nuc{Sn}{116}{} $ & 4.627 (1) & 356.5 & -18.41 & -0.3361 & 1.826 \\
    $ \nuc{Sn}{124}{} $ & 4.677 (1) & 352.5 & -18.24 & -0.3356 & 1.840 \\
    $ \nuc{Pb}{206}{} $ & 5.490 & 810.3 & -30.38 & -0.5527 & 1.820 \\
    $ \nuc{Pb}{208}{} $ & 5.503 (2) & 808.5 & -30.31 & -0.5524 & 1.823 \\ \hline \hline
  \end{tabular}
\end{table}
\begin{table}
  \centering
  \caption{Exchange Coulomb energies $ E_{\urm{x}} $ in the LDA and in the GGA by the B88, PW91, PBE, and PBEsol functionals. All units are in $ \mathrm{MeV} $\@.}
  \label{tab:gga_x}
  \begin{tabular}{rddddd}
    \hline \hline
    Nuclei & \multicolumn{1}{c}{LDA} & \multicolumn{1}{c}{B88} & \multicolumn{1}{c}{PW91} & \multicolumn{1}{c}{PBE} & \multicolumn{1}{c}{PBEsol} \\ \hline
    $ \nuc{He}{4}{} $ & -0.6494 & -0.7150 & -0.7290 & -0.7281 & -0.7030 \\
    $ \nuc{C}{12}{} $ & -1.962 & -2.077 & -2.109 & -2.105 & -2.056 \\
    $ \nuc{O}{16}{} $ & -2.638 & -2.773 & -2.812 & -2.806 & -2.748 \\
    $ \nuc{Ca}{40}{} $ & -7.087 & -7.319 & -7.395 & -7.381 & -7.277 \\
    $ \nuc{Ca}{48}{} $ & -7.113 & -7.349 & -7.420 & -7.409 & -7.305 \\
    $ \nuc{Ni}{58}{} $ & -10.28 & -10.57 & -10.66 & -10.65 & -10.52 \\
    $ \nuc{Sn}{116}{} $ & -18.41 & -18.81 & -18.94 & -18.92 & -18.74 \\
    $ \nuc{Sn}{124}{} $ & -18.24 & -18.64 & -18.77 & -18.75 & -18.57 \\
    $ \nuc{Pb}{206}{} $ & -30.38 & -30.91 & -31.09 & -31.06 & -30.81 \\
    $ \nuc{Pb}{208}{} $ & -30.31 & -30.84 & -31.02 & -30.99 & -30.74 \\ \hline \hline
  \end{tabular}
\end{table}
\begin{figure}
  \centering
  \includegraphics[width=8cm]{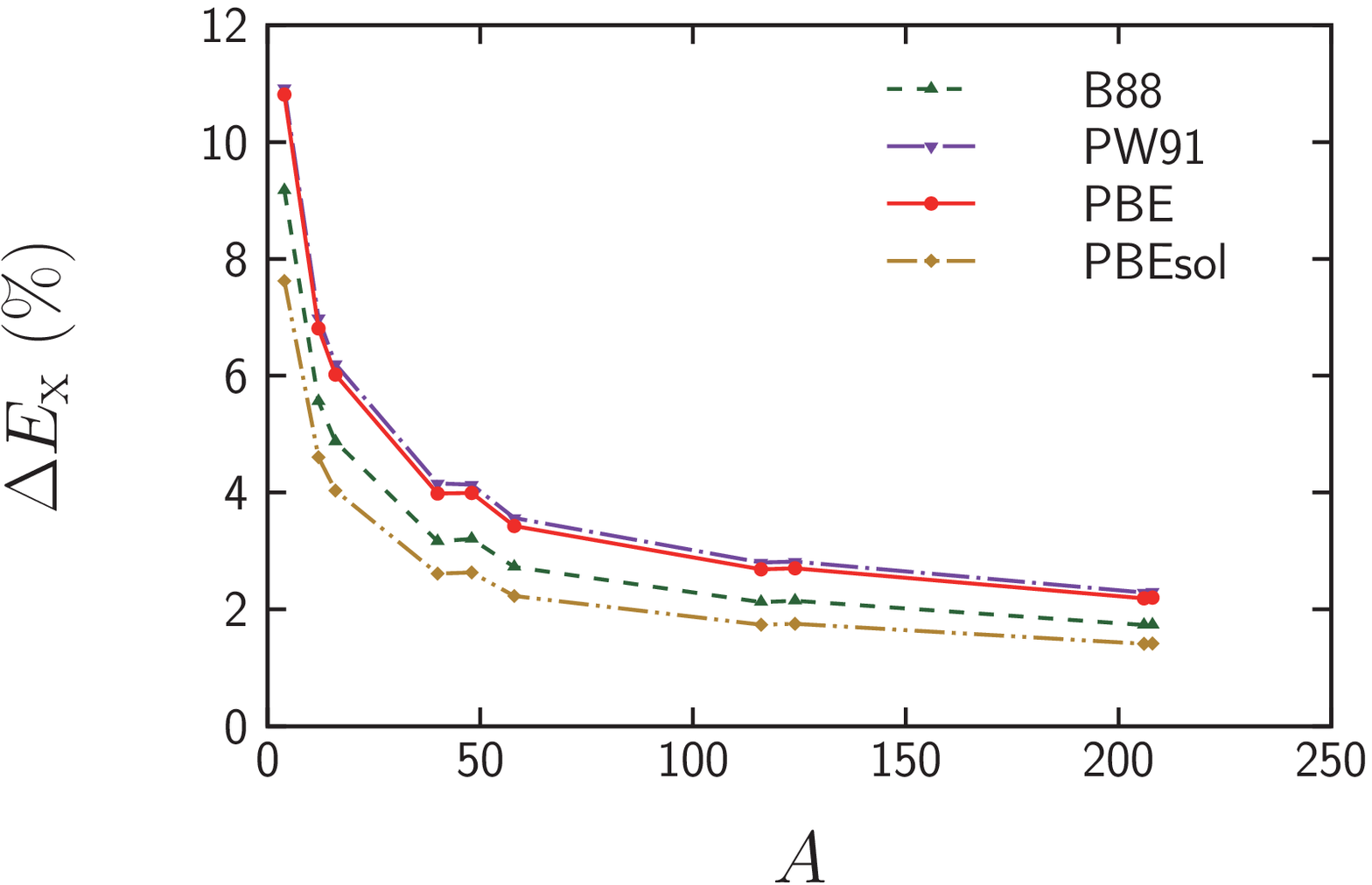}
  \caption{Deviation between the LDA and GGA in $ E_{\urm{x}} $ defined as Eq.~\eqref{eq:delta_x} as a function of $ A $.
    Those given by the B88, PW91, PBE, and PBEsol functionals are shown with the short-dashed, dot-dashed, solid, and dot-dot-dashed lines, respectively.}
  \label{fig:dev}
\end{figure}
\par
The direct, exchange, and correlation Coulomb energies in the LDA for selected nuclei from light to heavy regions are shown in Table~\ref{tab:lda}.
The experimental root-mean-square charge radii $ \left\langle r_{\urm{ch}}^2 \right\rangle^{1/2} $ \cite{Vries1987At.DataNucl.DataTables36_495} are also shown.
The corresponding exchange Coulomb energies for each functional in the GGA are shown in Table~\ref{tab:gga_x}.
\par
For isotopes, it is seen in Table~\ref{tab:lda} that the nucleus with larger radius gives smaller exchange and correlation energies.
The correlation energy is in general around $ 2 \, \% $ of the exchange energy, which is consistent with the estimate of order of magnitude in Fig.~\ref{fig:ratio}.
For electron systems, the correlation energy is around $ 10 \, \% $ of the exchange energy;
thus the correlation energy reduces the error of the Hartree-Fock-Slater approximation, i.e., the exchange LDA,
and thus the exchange-correlation LDA calculations often give reasonable results \cite{Engel}.
In contrast, for atomic nuclei, the correlation energy is only around $ 2 \, \% $ of the exchange energy, thus it does not change substantially the result of the Hartree-Fock-Slater approximation.
\par
In the whole mass region, the behaviors of the GGA functionals are similar to those in $ \nuc{Pb}{208}{} $.
The enhancements from LDA to GGA in light nuclei are larger than those in heavy nuclei because of the ratio of the surface region.
The deviations between the LDA and GGA in the exchange energy $ \Delta E_{\urm{x}} $ is shown in Fig.~\ref{fig:dev}.
\par
On the one hand, these non-empirical GGA functionals are determined to satisfy some conditions about the exchange-correlation hole for electron systems \cite{Perdew1996Phys.Rev.B54_16533},
but the exchange-correlation hole in atomic nuclei is determined mainly by the nuclear force.
Therefore, there is no guarantee to apply electron EDFs for nuclei, as discussed in Sec.~\ref{sec:electron_sep}.
\par
On the other hand, the deviation between the LDA exchange energy and the exact Fock energy
by self-consistent calculations was given by Le~Bloas \textit{et al\/.} \cite{LeBloas2011Phys.Rev.C84_014310}.
The result in Fig.~\ref{fig:dev} here shows a quite similar behavior compared with their work.
This indicates that the application of the electron GGA for atomic nuclei gives at least almost the same accuracy as the exact Hartree-Fock calculation, i.e., these functionals are valid for atomic nuclei as well as electron systems.
\par
In the self-consistent calculations, all potentials derived from these GGA functionals are local, and thus the numerical cost of the DFT calculations is $ O \left( N^3 \right) $.
In contrast, the Fock potential is non-local; hence the numerical cost of the Hartree-Fock calculations is $ O \left( N^4 \right) $ \cite{Engel}.
Therefore, the self-consistent calculations for Coulomb energy with GGA EDFs would be of high accuracy and low numerical cost, compared with the less accurate Hartree-Fock-Slater calculations or more accurate but higher-numerical-cost exact Hartree-Fock calculations.
Along this direction, some progress concerning the localized form of Fock terms in nuclear covariant DFT has been carried out \cite{Liang2012Phys.Rev.C86_021302(R), Gu2013Phys.Rev.C87_041301(R), VanGiai2014Phys.Scr.89_054008}.
In addition, these GGA EDFs hold the possibility of evaluating the electromagnetic contribution of the binding energy from the experimental data directly.
\par
One may consider the applicability of Coulomb correlation functionals for nuclear systems.
The test results of the Coulomb correlation energies $ E_{\urm{c}} $ are shown in Table I from light to heavy nuclei.
It is seen that in these calculations $ E_{\urm{c}} $ are all around $ 2 \, \% $ of $ E_{\urm{x}} $.
However, Bulgac and Shaginyan \cite{Bulgac1996Nucl.Phys.A601_103,Bulgac1999Phys.Lett.B469_1} evaluated that in atomic nuclei
$ E_{\urm{c}}/E_{\urm{x}} $ would be around $ -40 \, \% $ to $ -20 \, \% $, instead of $ 2 \, \% $.
To understand this big difference in a simple picture,
we keep in mind that between protons there exist not only weak repulsive Coulomb but also strong attractive nuclear interactions.
Hypothetically, if there is only Coulomb interaction,
since correlation always further decreases the energy of the whole system,
we have the signs of the Hartree, exchange, and correlation energies as
$ E_{\urm{d}} > 0 $, $ E_{\urm{x}} < 0 $, and $ E_{\urm{c}} < 0 $, respectively,
i.e.,
$ E_{\urm{c}} $ has the same sign as $ E_{\urm{x}} $.
In reality, as discussed in Sec.~\ref{sec:electron},
the correlation energy density functional is not separable at all.
In Refs.~\cite{Bulgac1996Nucl.Phys.A601_103,Bulgac1999Phys.Lett.B469_1}
the correlation functionals are written in terms of the response functions,
and such response functions are determined by the total interaction, i.e., mainly by the attractive nuclear part,
instead of the repulsive Coulomb part.
The total correlation energy is still negative,
which mainly comes from the contribution of the nuclear interaction.
As a result,
the contribution of the Coulomb interaction becomes positive;
i.e., for a Coulomb energies $ E_{\urm{c}} $ has a different sign than $ E_{\urm{x}} $.
In short, the correlation energy density functionals of electron systems cannot be applied directly to atomic nuclei.

%
%
%
%
\section{Conclusion and Perspectives}
\label{sec:conc}
\par
In this paper, we applied the exchange and correlation EDFs of the local density approximation and generalized gradient approximation in electron systems for atomic nuclei.
\par
For the exchange Coulomb energies, it is found that the deviation between the LDA and GGA $ \Delta E_{\urm{x}} $ ranges from around  $ 11 \, \% $ in $ \nuc{He}{4}{} $ to around $ 2.2 \, \% $ in $ \nuc{Pb}{208}{} $, by taking the PBE functional as an example of the GGA\@.
From light to heavy nuclei, it is seen that $ \Delta E_{\urm{x}} $ behaves in a very similar way as the deviation between the Hartree-Fock-Slater approximation and the exact Hartree-Fock calculation given by Le~Bloas \textit{et al\/.}~\cite{LeBloas2011Phys.Rev.C84_014310}.
In this sense, the GGA exchange functionals of electron systems are valid for atomic nuclei.
Furthermore, the numerical cost of GGA is $ O \left( N^3 \right) $, whereas that cost of the exact Hartree-Fock calculation is $ O \left( N^4 \right) $ for self-consistent calculations.
\par
In contrast, the correlation Coulomb energy density functionals of electron systems are not applicable for atomic nuclei,
because these functionals are not separable and the nuclear interaction determines the properties of atomic nuclei mainly.
\par
For future studies, we would like to use these Coulomb GGA functionals for self-consistent calculations.
There are two main open questions here.
One is the double counting of the correlation effects as we discussed in Sec.~\ref{sec:electron_sep}.
Another important point is the finite-size effect of protons, which electron systems do not suffer.
\par
So far, in most if not all of the DFT or Hartree-Fock calculations in nuclear physics the proton is treated as a point particle, and the Coulomb energy is calculated with the proton-density distribution $ \rho_p \left( r \right) $.
However, it is well known that the charge-density distribution is different from the point like proton-density distribution in atomic nuclei.
From the point of view of the electromagnetic force, the Coulomb energy should be calculated with the charge-density distribution. For example, it is given by the convolution of $ \rho_p \left( r \right) $ with the proton form factor as  \cite{Campi1972Nucl.Phys.A194_401}
\begin{equation}
  \rho_{\urm{ch}}
  \left( r \right)
  =
  \frac{1}{2 \pi^2 r}
  \int_0^{\infty}
  k \sin \left( k r \right) \,
  \tilde{\rho}_p \left( k \right) \,
  \exp
  \left[
    \frac{k^2}{4}
    \left( B^2 - a^2 \right)
  \right] \, dk,
\end{equation}
where
\begin{equation}
  B
  =
  \left(
    \frac{41.47}{A \hbar \omega}
  \right)^{1/2},
  \qquad
  a
  =
  \sqrt{\frac{2}{3}}
  \left\langle{r_p} \right\rangle,
  \qquad
  \left\langle{r_p} \right\rangle
  =
  0.8
  \, \mathrm{fm},
\end{equation}
$ A $ is the mass number,
$ \hbar\omega = 41 A^{-1/3}\,\mathrm{MeV}$ \cite{RingSchuck},
and $ \tilde{\rho}_p \left( k \right) $ is the Fourier transformation of $ \rho_p \left( r \right) $.
This leads to a general question about how to construct a DFT for the particles with finite size, which also corresponds to some progress in the DFT with frozen core approximation in condensed matter physics \cite{Parr1984J.Am.Chem.Soc.106_4049,Vargas2005J.Phys.Chem.A109_8880,Kiejna2006Phys.Rev.B73_035404,Kahros2013J.Chem.Phys.138_054110}.
%
%
%
\begin{acknowledgments}
\par
The authors appreciate Shinji Tsuneyuki for stimulating discussions and valuable comments.
The work was partially supported by the RIKEN iTHES project and iTHEMS program,
and the JSPS-NSFC Bilateral Program for Joint Research Project on Nuclear mass and life for unravelling mysteries of the r-process.
\end{acknowledgments}

%
%
%
\appendix
\section{Details of Coulomb Energy Density Functionals}
\label{sec:app_edf}
\subsection{Local Density Approximation}
\par
When it is assumed the energy density depends only on the density at $ \ve{r} $ locally as
\begin{equation}
  E_i \left[ \rho \right]
  =
  \int
  \epsilon_i^{\urm{LDA}} \left( \rho \left( \ve{r} \right) \right) \,
  \rho \left( \ve{r} \right) \,
  d \ve{r}
  \qquad
  \text{($ i = \mathrm{x}, \, \mathrm{c} $)},
\end{equation}
this approximation is called the LDA\@.
The LDA gives the exact solutions for the systems with homogeneous density distribution,
and it also gives high-accuracy results for the systems with nearly constant density distribution.
\par
In the homogeneous electron gas, the exchange energy density $ \epsilon_{\urm{x}} $ is known exactly.
That is used for the LDA exchange energy density $ \epsilon_{\urm{x}}^{\urm{LDA}} $.
For electron systems,
\begin{equation}
  \label{eq:LDA_x}
  \epsilon_{\urm{x}}^{\urm{LDA}}
  \left( \rho \right)
  =
  - \frac{3}{4}
  \left(
    \frac{3}{\pi}
  \right)^{1/3}
  \rho^{1/3}
  =
  - \frac{3}{4}
  \left(
    \frac{9}{4 \pi^2}
  \right)^{1/3}
  \frac{1}{r_{\urm{s}}},
\end{equation}
in the Hartree atomic unit, i.e., the electron mass $ m_e = 1 $, electron charge $ e^2 = 1 $, and $ 4 \pi \epsilon_0 = 1 $, while $ \hbar = 1 $ and $ c= 1/\alpha \simeq 137 $.
This is nowadays widely known as the Slater approximation \cite{Slater1951Phys.Rev.81_385}, derived by Dirac \cite{Dirac1930Proc.Camb.Phil.Soc.26_376}.
Here, $ r_{\urm{s}} $ is the Wigner-Seitz radius,
\begin{equation}
  r_{\urm{s}}
  =
  \left(
    \frac{3}{4 \pi \rho}
  \right)^{1/3}.
\end{equation}
\par
In contrast, the correlation energy density $ \epsilon_{\urm{c}} $ for the homogeneous electron gas is not known analytically.
In the LDA, it was derived by fitting for the ground-state energy of the homogeneous electron gas evaluated by the
diffusion Monte Carlo calculation \cite{Ceperley1980Phys.Rev.Lett.45_566}.
Several fittings of $ \epsilon_{\urm{c}} $ have been
proposed \cite{Vosko1980Can.J.Phys.58_1200--1211,Perdew1981Phys.Rev.B23_5048--5079,Perdew1992Phys.Rev.B45_13244--13249}.
One of the most widely used forms is PZ81 \cite{Perdew1981Phys.Rev.B23_5048--5079}, which reads
\begin{equation}
  \label{eq:LDA_c}
  \epsilon_{\urm{c}}^{\urm{PZ81}}
  \left( r_{\urm{s}} \right)
  =
  \begin{cases}
    -0.0480 + 0.0311 \ln r_{\urm{s}}
    - 0.0116 r_{\urm{s}} + 0.0020 r_{\urm{s}} \ln r_{\urm{s}}
    &
    \text{($ r_{\urm{s}} < 1 $)}, \\
    -0.1423 / \left(1 + 1.0529 \sqrt{r_{\urm{s}}} + 0.3334 r_{\urm{s}} \right)
    &
    \text{($ r_{\urm{s}} > 1 $)}.
  \end{cases}
\end{equation}
The LDA correlation function satisfies $ \epsilon_{\urm{c}}^{\urm{LDA}} \to -0.0480 + 0.0311 \ln r_{\urm{s}} $ in the high-density limit $ r_{\urm{s}} \to 0 $ \cite{Gell-Mann1957Phys.Rev.106_364}, which is satisfied by the PZ81 functional.
\subsection{Generalized Gradient Approximation}
\par
The DFT with LDA does not always represent correct results, which can be improved by the DFT with GGA (see, e.g.,~Ref.~\cite{Asada1992Phys.Rev.B46_13599}).
In the GGA, the energy density depends not only on the density distribution $ \rho $ but also on its gradient $ \left| \nabla \rho \right| $ at $ \ve{r} $ locally.
It is expressed as
\begin{equation}
  E_i \left[ \rho \right]
  =
  \int
  \epsilon_i^{\urm{GGA}} \left( \rho \left( \ve{r} \right), \left| \nabla \rho \left( \ve{r} \right) \right| \right) \,
  \rho \left( \ve{r} \right) \,
  d \ve{r}
  \qquad
  \text{($ i = \mathrm{x}, \, \mathrm{c} $)}.
\end{equation}
\par
Several non-empirical GGA functionals have been proposed \cite{Becke1988Phys.Rev.A38_3098, Perdew1992Phys.Rev.B46_6671, Perdew1996Phys.Rev.Lett.77_3865, Perdew2008Phys.Rev.Lett.100_136406}.
Most GGA exchange energy densities $ \epsilon_{\urm{x}}^{\urm{GGA}} $ are written as the product of the LDA counterpart $ \epsilon_{\urm{x}}^{\urm{LDA}} $ and an enhancement factor $ F^{\urm{GGA}} $:
\begin{equation}
  \label{eq:GGA_x}
  \epsilon_{\urm{x}}^{\urm{GGA}}
  \left( \rho, \left| \nabla \rho \right| \right)
  =
  \epsilon_{\urm{x}}^{\urm{LDA}}
  \left( \rho \right) \,
  F^{\urm{GGA}} \left( s \right).
\end{equation}
The enhancement factors of
the GGA-B88 \cite{Becke1988Phys.Rev.A38_3098}, GGA-PW91 \cite{Perdew1992Phys.Rev.B46_6671}, GGA-PBE \cite{Perdew1996Phys.Rev.Lett.77_3865} and
GGA-PBEsol \cite{Perdew2008Phys.Rev.Lett.100_136406} functionals are given below:
\begin{align}
  F^{\urm{B88}} \left( s \right)
  & =
    1
    +
    \frac{0.0168}{3}
    \left(
    \frac{\pi}{6}
    \right)^{1/3}
    \frac{\left[ 2 \left( 3 \pi^2 \right)^{1/3} s\right]^2}
    {1 + 0.0252 \left[ 2 \left( 3 \pi^2 \right)^{1/3} s\right] \sinh^{-1} \left[ 2 \left( 3 \pi^2 \right)^{1/3} s\right]}, \\
  F^{\urm{PW91}} \left( s \right)
  & =
    \frac{
    1 + 0.19645 s \sinh^{-1} \left( 7.7956 s \right)
    +
    \left( 0.2743 - 0.1508 e^{-100s^2} \right) s^2}
    {1 + 0.19645 s \sinh^{-1} \left( 7.7956 s \right) + 0.004s^4}  , \\
  F^{\urm{PBE}} \left( s \right)
  & =
    1 + 0.804
    -
    \frac{0.804}{1 + 0.21951 s^2/0.804} , \\
  F^{\urm{PBEsol}} \left( s \right)
  & =
    1 + 0.804
    -
    \frac{0.804}{1 + 0.1235 s^2/0.804}.
\end{align}
\subsection{Translation from Hartree Atomic Unit to General Unit}
\par
When we apply the EDFs of electron systems for atomic nuclei, we have to pay special attention to the relation between the Hartree atomic unit and the natural unit used in nuclear physics.
In the Hartree atomic unit, the energy and length units are
\begin{align}
  1 \, \mathrm{Hartree}
  & =
    \frac{\hbar^2}{m_e a_{\urm{B}}^2}, \\
  1 \, \text{$ \mathrm{a.u.} $ (length)}
  & =
    a_{\urm{B}},
\end{align}
respectively, where $ a_{\urm{B}} $ is the Bohr radius
\begin{equation}
  a_{\urm{B}}
  =
  \frac{4 \pi \epsilon_0 \hbar^2}{m_e e^2} = \frac{\hbar}{\alpha m_e c}.
\end{equation}
\par
When every quantity is written explicitly, the dimensionless Wigner-Seitz radius is given by
\begin{equation}
  r_{\urm{s}}
  =
  \left(
    \frac{3}{4 \pi \rho}
  \right)^{1/3}
  \frac{\alpha m_e c}{\hbar}.
\end{equation}
In order to create no confusion, we define a general dimensionless variable $ \xi $ as
\begin{equation}
  \xi
  =
  \left(
    \frac{3}{4 \pi \rho}
  \right)^{1/3}
  \frac{\alpha m c}{\hbar},
\end{equation}
where $ m $ is the corresponding mass of the particles, i.e., $ m_e $ in electron systems and $ m_p $ in atomic nuclei.
In addition, the energy unit now reads $ \alpha^2 m c^2 $.
With this general variable $ \xi $, the LDA exchange energy density in Eq.~\eqref{eq:LDA_x} reads
\begin{equation}
  \epsilon_{\urm{x}}^{\urm{LDA}}
  \left( \xi \right)
  =
  - \frac{3\alpha^2 m c^2}{4}
  \left(
    \frac{9}{4 \pi^2}
  \right)^{1/3}
  \frac{1}{\xi},
\end{equation}
and the LDA correlation energy density of PZ81 in Eq.~\eqref{eq:LDA_c} reads
\begin{equation}
  \epsilon_{\urm{c}}^{\urm{PZ81}}
  \left( \xi \right)
  =
  \begin{cases}
    \left( -0.0480 + 0.0311 \ln \xi
      - 0.0116 \xi + 0.0020 \xi \ln \xi
    \right) \alpha^2 m c^2
    &
    \text{($ \xi < 1 $)}, \\
    -0.1423 \alpha^2 m c^2/ \left(1 + 1.0529 \sqrt{\xi} + 0.3334 \xi \right)
    &
    \text{($ \xi > 1 $)}.
  \end{cases}
  \label{eq:LDA_genc}
\end{equation}
For the GGA, $ k_{\urm{F}} $ and $ s $ retain their forms as in Eqs.~\eqref{eq:GGA_s}.
\par
In the natural unit in nuclear physics, $ \hbar = c = 1 $ and
$ e^2 / 4\pi\epsilon_0 = \alpha \simeq 1/137 $ are used, and the units fm and MeV are connected via $ 1 = \hbar c \simeq 197.33 \, \mathrm{MeV} \, \mathrm{fm} $.
The proton mass is $ m_p \simeq 938.272 \, \mathrm{MeV} $\@.
\subsection{Order-of-Magnitude Estimates}
\label{sec:app_order}
\begin{figure}
  \centering
  \includegraphics[width=8cm]{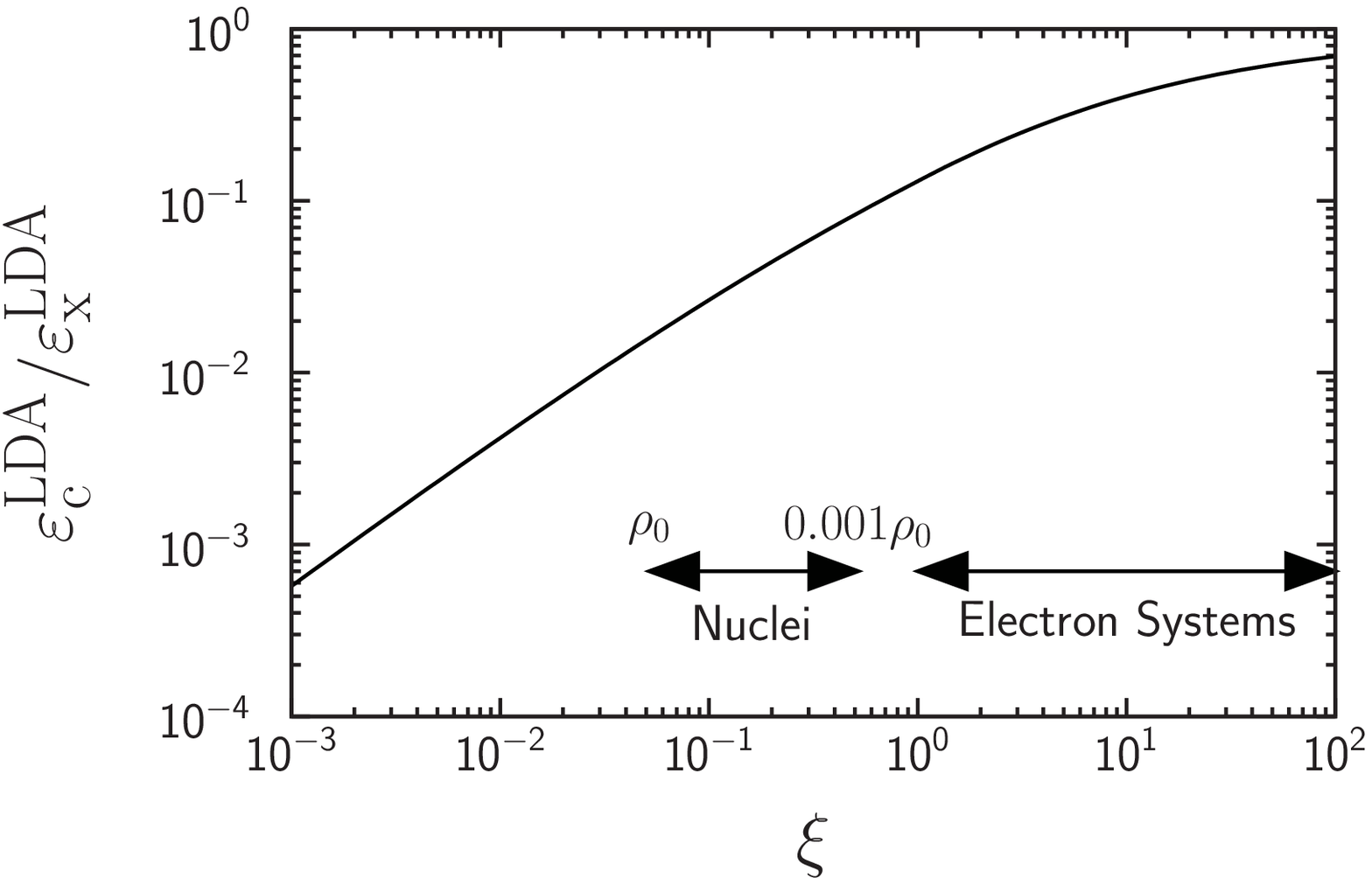}
  \caption{Ratio of the correlation energy density to the exchange energy density $ \epsilon_{\urm{c}} / \epsilon_{\urm{x}} $ in LDA with the PZ81 \cite{Perdew1981Phys.Rev.B23_5048--5079} functional as a function of $ \xi $.}
  \label{fig:ratio}
\end{figure}
\par
In Fig.~\ref{fig:ratio}, the ratio of the correlation energy density to the exchange energy density $ \epsilon_{\urm{c}} / \epsilon_{\urm{x}} $ in LDA is shown as a function of $ \xi $.
For electron systems, the range of $ \xi $ is generally $ 1 \lesssim \xi \lesssim 100 $.
It is seen in the figure that the correlation energy $ E_{\urm{c}} $ is of $ O \left( 10^{-1} \right) $ with respect to the exchange energy $ E_{\urm{x}} $.
In the limit of $ \xi \to \infty $,
$ \epsilon_{\urm{c}} / \epsilon_{\urm{x}} $ goes to $ 0.9316 $.
\par
In contrast, the nuclear saturation density $ \rho_0 \simeq 0.16 \, \mathrm{fm}^{-3} $ corresponds to $ \xi_0 \simeq 0.052 $.
When the density $ \rho $ drops by three orders of magnitude, the corresponding $ \xi $ increases by one order of magnitude.
This range is illustrated in Fig.~\ref{fig:ratio}.
Therefore, $ E_{\urm{c}}/E_{\urm{x}} $ is of $ O \left( 10^{-2} \right) $ in atomic nuclei.
In the limit of $ \xi \to 0 $, $ \epsilon_{\urm{c}} / \epsilon_{\urm{x}} $ goes to zero.
\par
In terms of the fine-structure constant $ \alpha $, the exchange energy density $ \epsilon_{\urm{x}} $ in LDA is exactly proportional to $ \alpha $; i.e., the exchange energy comes from the two-body Coulomb interaction only.
For the correlation energy density $ \epsilon_{\urm{c}} $ in LDA, it is found that $ \epsilon_{\urm{c}} $ is also proportional to $ \alpha $ in the case of large $ \xi $.
This indicates at the low-density limit, e.g., in electron systems, the leading-order contribution to the correlation energy also comes from the two-body Coulomb interaction.
In contrast, in the case of small $ \xi $, $ \epsilon_{\urm{c}} $ is of the order of $ O \left( \alpha^2\log\alpha \right) $.
This implies in atomic nuclei the leading-order contribution to the correlation energy comes from beyond the two-body interaction.
\begin{figure}
  \centering
  \includegraphics[width=8cm]{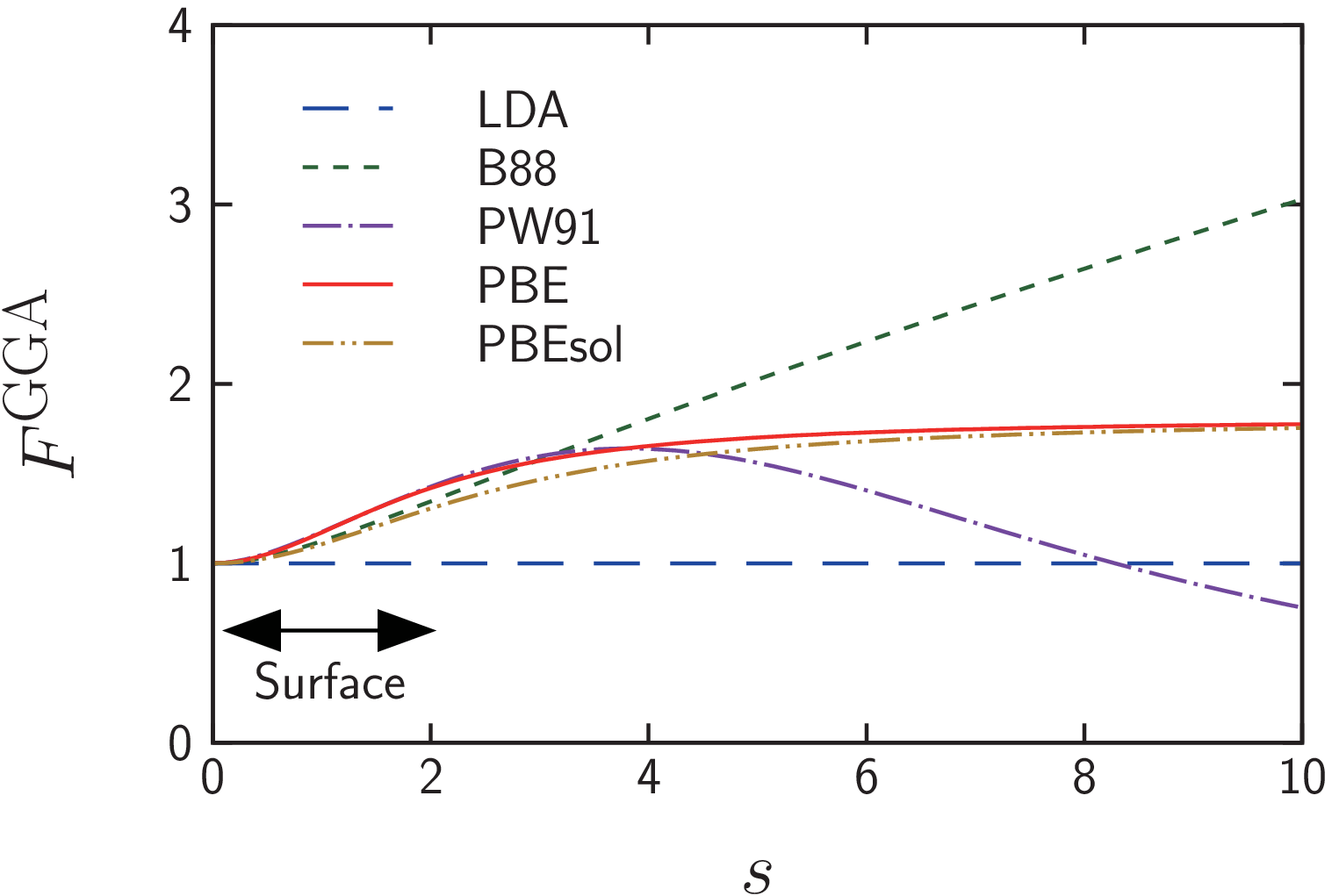}
  \caption{GGA exchange enhancement factors $ F^{\urm{GGA}} $ as a function of $ s $.
    Those given by the B88 \cite{Becke1988Phys.Rev.A38_3098}, PW91 \cite{Perdew1992Phys.Rev.B46_6671}, PBE \cite{Perdew1996Phys.Rev.Lett.77_3865}, and PBEsol \cite{Perdew2008Phys.Rev.Lett.100_136406} functionals are shown with the short-dashed, dot-dashed, solid, and dot-dot-dashed lines, respectively.
    For comparison, $ F \equiv 1 $ by the LDA is shown with the long-dashed line.}
  \label{fig:gga-f}
\end{figure}
\par
In Fig.~\ref{fig:gga-f}, the GGA exchange enhancement factors $ F^{\urm{GGA}} $ given by the B88, PW91, PBE, and PBEsol functionals are shown as a function of $ s $.
It is seen that all four GGA functionals behave similarly in the range $ 0 \lesssim s \lesssim 3 $, before they start to diverge from each other.
In this region, $ F\left(s\right) \ge 1 $, which means the absolute value of the GGA exchange energy is larger than that of the LDA\@.
%
%
\bibliography{proton_gga}
\end{document}